
\def\singlespace{\normalbaselines}
\def\oneandahalfspace{\baselineskip=1.15\normalbaselineskip plus 1pt
\lineskip=2pt\lineskiplimit=1pt}

\def\np{\vfill\eject}
\def\nl{\hfil\break}

\def\nofirstpagenoten{\nopagenumbers\footline={\ifnum\pageno>1\tenrm
\hss\folio\hss\fi}}
\def\nofirstpagenotwelve{\nopagenumbers\footline={\ifnum\pageno>1\twelverm
\hss\folio\hss\fi}}
\def\leaderfill{\leaders\hbox to 1em{\hss.\hss}\hfill}
\def\ft#1#2{{\textstyle{{#1}\over{#2}}}}
\def\frac#1/#2{\leavevmode\kern.1em
\raise.5ex\hbox{\the\scriptfont0 #1}\kern-.1em/\kern-.15em
\lower.25ex\hbox{\the\scriptfont0 #2}}
\def\sfrac#1/#2{\leavevmode\kern.1em
\raise.5ex\hbox{\the\scriptscriptfont0 #1}\kern-.1em/\kern-.15em
\lower.25ex\hbox{\the\scriptscriptfont0 #2}}


\parindent=20pt
\def\narrow{\advance\leftskip by 40pt \advance\rightskip by 40pt}

\def\AB{\bigskip
        \centerline{\bf ABSTRACT}\medskip\narrow}
\def\nonarrower{\advance\leftskip by -40pt\advance\rightskip by -40pt}
\def\AE{\bigskip\nonarrower}

\def\boxit#1{\vbox{\hrule\hbox{\vrule\kern3pt
        \vbox{\kern3pt#1\kern3pt}\kern3pt\vrule}\hrule}}

\def\gtorder{\mathrel{\raise.3ex\hbox{$>$}\mkern-14mu
             \lower0.6ex\hbox{$\sim$}}}
\def\ltorder{\mathrel{\raise.3ex\hbox{$<$}|mkern-14mu
             \lower0.6ex\hbox{\sim$}}}
\def\dalemb#1#2{{\vbox{\hrule height .#2pt
        \hbox{\vrule width.#2pt height#1pt \kern#1pt
                \vrule width.#2pt}
        \hrule height.#2pt}}}

\font\fourteentt=cmtt10 scaled \magstep2
\font\fourteenbf=cmbx12 scaled \magstep1
\font\fourteenrm=cmr12 scaled \magstep1
\font\fourteeni=cmmi12 scaled \magstep1
\font\fourteenss=cmss12 scaled \magstep1
\font\fourteensy=cmsy10 scaled \magstep2
\font\fourteensl=cmsl12 scaled \magstep1
\font\fourteenex=cmex10 scaled \magstep2
\font\fourteenit=cmti12 scaled \magstep1
\font\twelvett=cmtt10 scaled \magstep1 \font\twelvebf=cmbx12
\font\twelverm=cmr12 \font\twelvei=cmmi12
\font\twelvess=cmss12 \font\twelvesy=cmsy10 scaled \magstep1
\font\twelvesl=cmsl12 \font\twelveex=cmex10 scaled \magstep1
\font\twelveit=cmti12
\font\tenss=cmss10
 
 \font\ninebf=cmbx7 scaled \magstep1
\font\ninerm=cmr7 scaled \magstep1 \font\ninei=cmmi7 scaled \magstep1
\font\ninesy=cmsy7 scaled \magstep1 
\font\eightrm=cmr7 scaled 1140 
 
\font\sevenbf=cmbx7 \font\sevenrm=cmr7 \font\seveni=cmmi7
\font\sevensy=cmsy7 

\catcode`@=11
\newskip\ttglue
\newfam\ssfam

\def\fourteenpoint{\def\rm{\fam0\fourteenrm}
\textfont0=\fourteenrm \scriptfont0=\tenrm \scriptscriptfont0=\sevenrm
\textfont1=\fourteeni \scriptfont1=\teni \scriptscriptfont1=\seveni
\textfont2=\fourteensy \scriptfont2=\tensy \scriptscriptfont2=\sevensy
\textfont3=\fourteenex \scriptfont3=\fourteenex \scriptscriptfont3=\fourteenex
\def\it{\fam\itfam\fourteenit} \textfont\itfam=\fourteenit
\def\sl{\fam\slfam\fourteensl} \textfont\slfam=\fourteensl
\def\bf{\fam\bffam\fourteenbf} \textfont\bffam=\fourteenbf
\scriptfont\bffam=\tenbf \scriptscriptfont\bffam=\sevenbf
\def\tt{\fam\ttfam\fourteentt} \textfont\ttfam=\fourteentt
\def\ss{\fam\ssfam\fourteenss} \textfont\ssfam=\fourteenss
\tt \ttglue=.5em plus .25em minus .15em
\normalbaselineskip=16pt
\abovedisplayskip=16pt plus 4pt minus 12pt
\belowdisplayskip=16pt plus 4pt minus 12pt
\abovedisplayshortskip=0pt plus 4pt
\belowdisplayshortskip=9pt plus 4pt minus 6pt
\parskip=5pt plus 1.5pt
\setbox\strutbox=\hbox{\vrule height12pt depth5pt width0pt}
\let\sc=\tenrm
\let\big=\fourteenbig \normalbaselines\rm}
\def\fourteenbig#1{{\hbox{$\left#1\vbox to12pt{}\right.\n@space$}}}

\def\twelvepoint{\def\rm{\fam0\twelverm}
\textfont0=\twelverm \scriptfont0=\ninerm \scriptscriptfont0=\sevenrm
\textfont1=\twelvei \scriptfont1=\ninei \scriptscriptfont1=\seveni
\textfont2=\twelvesy \scriptfont2=\ninesy \scriptscriptfont2=\sevensy
\textfont3=\twelveex \scriptfont3=\twelveex \scriptscriptfont3=\twelveex
\def\it{\fam\itfam\twelveit} \textfont\itfam=\twelveit
\def\sl{\fam\slfam\twelvesl} \textfont\slfam=\twelvesl
\def\bf{\fam\bffam\twelvebf} \textfont\bffam=\twelvebf
\scriptfont\bffam=\ninebf \scriptscriptfont\bffam=\sevenbf
\def\tt{\fam\ttfam\twelvett} \textfont\ttfam=\twelvett
\def\ss{\fam\ssfam\twelvess} \textfont\ssfam=\twelvess
\tt \ttglue=.5em plus .25em minus .15em
\normalbaselineskip=14pt
\abovedisplayskip=14pt plus 3pt minus 10pt
\belowdisplayskip=14pt plus 3pt minus 10pt
\abovedisplayshortskip=0pt plus 3pt
\belowdisplayshortskip=8pt plus 3pt minus 5pt
\parskip=3pt plus 1.5pt
\setbox\strutbox=\hbox{\vrule height10pt depth4pt width0pt}
\let\sc=\ninerm
\let\big=\twelvebig \normalbaselines\rm}
\def\twelvebig#1{{\hbox{$\left#1\vbox to10pt{}\right.\n@space$}}}

\def\tenpoint{\def\rm{\fam0\tenrm}
\textfont0=\tenrm \scriptfont0=\sevenrm \scriptscriptfont0=\fiverm
\textfont1=\teni \scriptfont1=\seveni \scriptscriptfont1=\fivei
\textfont2=\tensy \scriptfont2=\sevensy \scriptscriptfont2=\fivesy
\textfont3=\tenex \scriptfont3=\tenex \scriptscriptfont3=\tenex
\def\it{\fam\itfam\tenit} \textfont\itfam=\tenit
\def\sl{\fam\slfam\tensl} \textfont\slfam=\tensl
\def\bf{\fam\bffam\tenbf} \textfont\bffam=\tenbf
\scriptfont\bffam=\sevenbf \scriptscriptfont\bffam=\fivebf
\def\tt{\fam\ttfam\tentt} \textfont\ttfam=\tentt
\def\ss{\fam\ssfam\tenss} \textfont\ssfam=\tenss
\tt \ttglue=.5em plus .25em minus .15em
\normalbaselineskip=12pt
\abovedisplayskip=12pt plus 3pt minus 9pt
\belowdisplayskip=12pt plus 3pt minus 9pt
\abovedisplayshortskip=0pt plus 3pt
\belowdisplayshortskip=7pt plus 3pt minus 4pt
\parskip=0.0pt plus 1.0pt
\setbox\strutbox=\hbox{\vrule height8.5pt depth3.5pt width0pt}
\let\sc=\eightrm
\let\big=\tenbig \normalbaselines\rm}
\def\tenbig#1{{\hbox{$\left#1\vbox to8.5pt{}\right.\n@space$}}}
\let\rawfootnote=\footnote \def\footnote#1#2{{\rm\parskip=0pt\rawfootnote{#1}
{#2\hfill\vrule height 0pt depth 6pt width 0pt}}}

\def\tenfoot{\tenpoint\hskip-\parindent\hskip-.1cm}

\overfullrule=0pt
\twelvepoint
\def\sbullet{\raise.2em\hbox{$\scriptscriptstyle\bullet$}}
\nofirstpagenotwelve
\hsize=16.5 truecm
\baselineskip 15pt

\def\ft#1#2{{\textstyle{{#1}\over{#2}}}}

\def\ket#1{\big| #1\big\rangle}

\def\dum{{\phantom{X}}}

\def\phys{\big|{\rm phys}\big\rangle}

\def\del{\partial}

\def\phys{\big|\hbox{phys}\big\rangle}

\oneandahalfspace
\rightline{CTP TAMU--22/93}
\rightline{hep-th/9304115}
\rightline{April 1993}

\vskip 2truecm
\centerline{\bf On Higher-spin Generalisations of String Theory}
\vskip 1.5truecm
\centerline{H. Lu, C.N. Pope,\footnote{$^*$}{\tenfoot Supported in part
by the U.S. Department of Energy, under
grant DE-FG05-91ER40633.}
and X.J.
Wang\footnote{}{\tenfoot }}
\vskip 1.5truecm
\centerline{\it Center
for Theoretical Physics,
Texas A\&M University,}
\centerline{\it College Station, TX 77843--4242, USA.}

\vskip 1.5truecm
\AB\singlespace
    We construct BRST operators for certain higher-spin extensions of
the Virasoro algebra, in which there is a spin-$s$ gauge field on the
world sheet, as well as the spin-2 gauge field corrresponding to the
two-dimensional metric.  We use these BRST operators to study the physical
states of the associated string theories, and show how they are related to
certain minimal models.
\AE\oneandahalfspace

\np
\noindent
{\bf 1. Introduction}
\bigskip

    Two-dimensional gauge theories are the progenitors of string theories.
By gauging the Virasoro algebra, realised as a semi-local symmetry of a
set of free scalar fields, one is led to the usual bosonic string.  If
one gauges the supersymmetric extension of the Virasoro algebra, one
obtains a supersymmetric string theory.  Since there exist many other
extensions of the Virasoro algebra, in which one adds currents with spins
greater than two, it is natural to enquire whether these too can give rise to
interesting generalisations of string theory.  Such algebras are
generically called $W$ algebras, and the resulting string theories are
called $W$ strings.  A general feature of the $W$ algebras is that they
are non-linear, and this leads to certain complications both of a
conceptual and a technical nature when one builds the corresponding
string theories.

     The simplest example of a $W$ algebra is the $W_3$ algebra of
Zamolodchikov [1].  This contains a spin-3 primary current $W$ in addition
to the energy-momentum tensor $T$.  The idea of gauging the algebra, to
obtain a two-dimensional theory of $W_3$ gravity, was first put forward
in [2].  This leads on naturally to the idea of building a $W_3$ string
[3,4,5]. The first requirement for building such a string is an anomaly-free
theory of $W_3$ gravity.  In [5], it was shown how such a theory may be
built, using standard BRST techniques.  The BRST operator $Q_B$ for the
$W_3$ algebra had been found in [6,7], where it was shown that nilpotency
demands that the matter currents $T$ and $W$ should generate the $W_3$
algebra with central charge $c=100$.  Multi-scalar realisations were found in
[8], in terms of a set of fields $(\varphi,X^\mu)$.  The scalar
$\varphi$ plays a special r\^ole, which we shall explain later, whilst
the $d$ scalars $X^\mu$ appear in $T$ and $W$ only {\it via} their
energy-momentum tensor $T^{\rm eff}$.  The fields $X^\mu$ will acquire the
interpretation of coordinates on an effective target spacetime.  For $T$ and
$W$ to generate the $W_3$ algebra with $c=100$, it is necessary for $T^{\rm
eff}$ to generate the Virasoro algebra with central charge $c^{\rm
eff}=\ft{51}2$ [4,5].  Thus a background charge is needed regardless of the
dimension $d$ of the spacetime described by $X^\mu$, and so there is no notion
of a ``critical dimension'' for the theory.

     The spectrum of physical states can be studied most elegantly in the
BRST formalism:  A physical state $\ket{\chi}$ is one that is
annihilated by the BRST operator but is not BRST trivial.  Thus, $Q_B
\ket{\chi}=0$ but $\ket{\chi}\ne Q_B\ket{\psi}$ for any state
$\ket{\psi}$.  It was shown recently that a dramatic simplification of
the BRST operator, and the physical states, can be achieved by performing
a redefinition under which the ghost fields $(b,c)$ for the spin-2
current and $(\beta,\gamma)$ for the spin-3 current are mixed with the
special scalar $\varphi$ mentioned above [9].  In terms of the redefined
fields, the BRST operator becomes [9]
$$
\eqalignno{
Q_B&= Q_0 + Q_1,&(1.1)\cr
Q_0&=\oint dz\, c \Big(T^{\rm eff} +T_{\varphi} + T_{\gamma,\beta} + \ft12
T_{c,b} \Big), &(1.2)\cr
Q_1&=\oint dz\, \gamma\Big( (\del\varphi)^3 + 3\alpha\, \del^2\varphi\, \del
\varphi + \ft{19}8 \del^3\varphi +\ft92 \del\varphi\, \beta\, \del\gamma
+\ft32 \alpha\, \del\beta\, \del\gamma\Big),&(1.3)\cr}
$$
where the energy-momentum tensors are given by
$$
\eqalignno{
T_\varphi&\equiv -\ft12 (\del\varphi)^2 -\alpha\, \del^2\varphi, &(1.4)\cr
T_{\gamma,\beta}&\equiv -3\, \beta\,\del\gamma -2\, \del\beta\, \gamma,
&(1.5)\cr
T_{c,b}&\equiv -2\, b\, \del c - \del b\, c, &(1.6)\cr
T^{\rm eff} &\equiv -\ft12 \del X^\mu\, \del X^\nu\, \eta_{\mu\nu} -
i a_\mu\, \del^2 X^\mu. &(1.7)\cr}
$$
The background charge $\alpha$ for the scalar $\varphi$ is given by
$\alpha^2=\ft{49}8$, and the background-charge vector $a_\mu$ is chosen so that
$d-12a_\mu \, a^\mu =\ft{51}2$.  The BRST operator is graded, with $Q_0^2=Q_1^2
=\{Q_0,Q_1\}=0$.

     It was found in [10,11,9] that the physical states of the $W_3$ string
comprise four basic sectors.  The first three sectors are described by
physical operators $V$ of the form [9]
$$
V_\Delta=c\, U(\beta,\gamma,\varphi)\, V^{\rm eff}(X),\eqno(1.8)
$$
where $[Q_0,V_\Delta \}=0$, and the $Q_1$ constraint reduces to
$[Q_1,U(\beta,\gamma,\varphi) \}=0$.  These conditions imply that $V^{\rm
eff}$ is an effective physical operator in the spacetime $X^\mu$,
corresponding to effective physical states $\phys_{\rm eff}\equiv V^{\rm
eff}(X(0))\ket{0}$ satisfying the highest-weight conditions
$$
\eqalign{
L^{\rm eff}_n\phys_{\rm eff} &=0,\qquad n>0,\cr
(L^{\rm eff}_0 -\Delta)\phys_{\rm eff}&=0.\cr}\eqno(1.9)
$$
The intercept $\Delta$ takes the values $\Delta=1,\, \ft{15}{16},\,
\ft12$ in the three sectors.  The fourth sector of physical states is
described by operators of the form [9]
$$
V_0=c\, U_1(\beta,\gamma,\varphi) + U_2(\beta,\gamma,\varphi).\eqno(1.10)
$$
These correspond to discrete states, with zero momentum in the effective
spacetime.  The operators $V_\Delta$ in the first three
sectors can be viewed as representatives of the physical operators for
effective Virasoro strings with the three intercept values $\Delta$ given
above, whilst all the discrete operators $V_0$ can be viewed as
representatives of the identity operator in the effective spacetime.
Many examples for all four sectors, up to level 9 in excitations of the
$(\varphi,\beta,\gamma)$ system, are given in [9].

     A procedure for computing scattering amplitudes for the physical
states of the $W_3$ string was presented in [11,9].  It was found by studying
many examples that the amplitudes are characterised by the weights
$\Delta$ of the physical operators $V_\Delta$ appearing in the correlation
functions. If one associates the operators $V_1$, $V_{15/16}$ and
$V_{1/2}$ respectively with the weight $\{0,\ft{15}{16},\ft12\}$ primary
fields $\{1,\sigma,\varepsilon\}$ of the Ising model, then the pattern
of vanishing and non-vanishing three-point functions is in one-to-one
correspondence with the fusion rules of the Ising model [11].  The
four-point and higher-point functions exhibit duality and factorisation
properties that are consistent with these underlying three-point
functions.  In fact the picture that emerges is that the $W_3$ string,
and its interactions, can really be viewed as being described by a
special case of an ordinary conformal field theory in which one
takes the tensor product of the ($c=\ft12$) Ising model with a $c=\ft{51}2$
energy-momentum tensor for $d$ scalar fields $X^\mu$ with a background
charge.  (See [9,12] for a discussion of a minor technicality associated with
the need to include the discrete operators $V_0$ as representatives of the
identity in correlation functions in order to reproduce the full set of
correlation functions for the Ising model.)  From the effective spacetime
point of view, the scattering amplitudes for the $W_3$ string, obtained
using the procedure introduced in [11,9], coincide with those for the tensor
product of the Ising model and a $c=\ft{51}2$ Virasoro string, which was
studied in [13] by using the sophisticated ``Group Theoretic Method.''

\bigskip
\noindent{\bf 2. Higher-spin generalisations}
\bigskip

     A natural generalisation of the $W_3$ string is to study the $W_N$
string, obtained by gauging the $W_N$ algebra.  This has currents of
spins $3,\ 4,\, \ldots,\, N$ in addition to the energy-momentum tensor
$T$.  Since the BRST operator for the $W_N$ algebra is not known, except
for the case of $W_3$, a complete investigation for general $N$ is not
possible.  However, by making certain plausible assumptions [4] the
spectrum of physical states with standard ghost structure has been
determined [14].  The indications are that the full spectrum of physical
states will coincide with those that would be obtained by taking the
tensor product of the $N$'th unitary Virasoro minimal model with an
energy-momentum tensor $T^{\rm eff}$, with central charge given by
$c=26-\Big[1-{6\over N(N+1)}\Big]$, for the target spacetime fields $X^\mu$.

     A simpler possibility for higher-spin generalisations of the $W_3$ string
is to consider the case of an algebra involving just two currents,
namely the energy-momentum tensor $T$ and a spin-$s$ current $W$.  The
case $s=3$ corresponds to the $W_3$ algebra itself.  It should be
possible to realise such algebras in terms of a scalar field $\varphi$
and an energy-momentum tensor $T^{\rm eff}$ for a set of scalars $X^\mu$,
analogous to the realisation of $W_3$.  One would expect
that the BRST operator for such an algebra could again be simplified
considerably by the kind of redefinitions amongst the ghost and
$\varphi$ field that were obtained for $W_3$ in [9], and which lead to
the form (1.1--1.3) for its BRST operator.  In fact we can take a short
cut to the construction of the BRST operator in the redefined
formalism by simply making an ansatz that appropriately generalises
(1.1--1.3) for the BRST operator, and then requiring that it be nilpotent.
Using the same notation as for the $W_3$ case, we introduce the usual
$(b,c)$ ghost sytem for the spin-2 current, and the $(\beta,\gamma)$
ghost sytem for the spin-$s$ current.  Note that $\beta$ therefore has
spin $s$, and $\gamma$ has spin $(1-s)$. Thus we begin by writing
$$
\eqalignno{ Q_B&= Q_0 + Q_1,&(2.1)\cr
Q_0&=\oint dz\, c \Big(T^{\rm eff} +T_{\varphi} + T_{\gamma,\beta} + \ft12
T_{c,b} \Big), &(2.2)\cr
Q_1&=\oint dz\, \gamma \, F(\beta,\gamma,\varphi),&(2.3)\cr}
$$
where the energy-momentum tensors are given by
$$
\eqalignno{
T_\varphi&\equiv -\ft12 (\del\varphi)^2 -\alpha\, \del^2\varphi, &(2.4)\cr
T_{\gamma,\beta}&\equiv -s\, \beta\,\del\gamma -(s-1)\, \del\beta\, \gamma,
&(2.5)\cr
T_{c,b}&\equiv -2\, b\, \del c - \del b\, c, &(2.6)\cr
T^{\rm eff} &\equiv -\ft12 \del X^\mu\, \del X^\nu\, \eta_{\mu\nu} -
i a_\mu\, \del^2 X^\mu. &(2.7)\cr}
$$
The operator $F(\beta,\gamma,\varphi)$ has spin $s$ and ghost number
zero, generalising the quantity appearing in (1.3) in the case of $W_3$.

     We expect on general grounds that operators $Q_0$ and $Q_1$ defined
as above should exist, satisfying $Q_0^2=Q_1^2=\{Q_0,Q_1\}=0$. One way to
see this is to note that one expects that a BRST operator for $W_N$, with
a grading $\widetilde Q_B=\sum_{i=0}^{N-2} \widetilde Q_i$,
$\{\widetilde Q_i,\widetilde Q_j\}=0$, should exist, generalising the grading
of the $W_3$ case discussed in section 1 [15].  A realisation of $W_N$ can be
given in terms of the currents of $W_{N-1}$ and one additional scalar,
$\varphi$ [8,4,14].  The  highest term in $\widetilde Q_B$, namely $\widetilde
Q_{N-2}$, would involve just the field $\varphi$ and the ghost system for the
spin-$N$ current of $W_N$ [16].  Thus the nilpotency of $\widetilde Q_{N-2}$
would be a property that is independent of the nature of the currents of lower
spin, and so the form of $Q_1$ in our BRST operator for the spin-2
plus spin-$s$ system will be the same as that of $\widetilde Q_{N-2}$ for
the BRST operator of $W_s$.  It was shown in [14] that the scalar $\varphi$
involved in the realisation of $W_s$ currents in terms of $W_{s-1}$
currents must have a background charge $\alpha$ given by
$$
\alpha^2={(s-1)(2s+1)^2\over 4(s+1)}\eqno(2.8)
$$
in order to achieve the correct central charge for criticality.  Thus
we expect that for each $s$ there should be at least a nilpotent $Q_B$
given by (2.1--2.3) with the parameter $\alpha$ given by (2.8).  In fact, as we
shall see, we find by demanding the nilpotency of $Q_B$ defined in (2.1)
that in addition to the above solution, with $\alpha$ given by (2.8), there
can be further solutions, with different operators $F(\beta,\gamma,\varphi)$
and different values of $\alpha$.  These seem to be unrelated to the $W_s$
algebra.

     Owing to the grading of $Q_B$, the conditions for its nilpotency
separate into the conditions $Q_0^2=0$, $\{Q_0,Q_1\}=0$ and $Q_1^2=0$.  The
first of these is easily analysed in general.  Nilpotency of $Q_0$ is
achieved provided that the total central charge vanishes, {\it i.e.}
$$
0=-26 -2(6s^2-6s+1) + 1+12\alpha^2 + c^{\rm eff}.\eqno(2.9)
$$
The effective central charge $c^{\rm eff}$ is given by $c^{\rm
eff}=d-12a_\mu \, a^\mu$.  For the remaining nilpotency conditions, we shall
carry out a case-by-case analysis for $s=4$, 5 and 6.

     We begin with the case $s=4$. The most general ansatz for the spin-4
ghost number zero operator $F(\beta,\gamma,\varphi)$ in (2.3) has twelve
terms, two of which can be eliminated since they correspond to total
derivatives.  The condition $\{Q_0,Q_1\}=0$ then leads to a system of linear
equations for the coefficients of the remaining terms, and the condition
$Q_1^2=0$ gives rise to quadratic equations for the coefficients.  We find two
distinct solutions; one with $\alpha^2=\ft{243}{20}$ which corresponds to the
expected $s=4$ case of (2.8) associated with the $W_4$ algebra, and the other
with $\alpha^2=\ft{361}{30}$.  For the first solution, with
$\alpha^2=\ft{243}{20}$, we find
$$
\eqalign{
F(\beta,\gamma,\varphi)&=(\del\varphi)^4 + 4\alpha\, \del^2\varphi\,
(\del\varphi)^2 + \ft{41}5 (\del^2\varphi)^2 + \ft{124}{15}
\del^3\varphi\, \del \varphi +\ft{46}{135} \alpha\, \del^4\varphi\cr
& +8 (\del\varphi)^2\, \beta\, \del\gamma -\ft{16}9\alpha\, \del^2\varphi\,
\beta\, \del\gamma -\ft{32}9 \alpha\, \del\varphi\,
\beta\, \del^2 \gamma-\ft45 \beta\, \del^3\gamma + \ft{16}3 \del^2\beta\,
\del\gamma .\cr}\eqno(2.10)
$$
For the second solution, with $\alpha^2=\ft{361}{30}$, we find the
corresponding operator $F(\beta,\gamma,\varphi)$ in (2.3) is given by
$$
\eqalign{
F(\beta,\gamma,\varphi)&=(\del\varphi)^4 + 4\alpha\, \del^2\varphi\,
(\del\varphi)^2 + \ft{253}{30} (\del^2\varphi)^2 + \ft{39}{5}
\del^3\varphi\, \del \varphi +\ft{41}{570} \alpha\, \del^4\varphi\cr
& +8 (\del\varphi)^2\, \beta\, \del\gamma -\ft{26}{19}\alpha\, \del^2\varphi\,
\beta\, \del\gamma -\ft{66}{19} \alpha\, \del\varphi\,
\beta\, \del^2 \gamma-\ft{26}{15} \beta\, \del^3\gamma + \ft{29}5
\del^2\beta\, \del\gamma .\cr}\eqno(2.11)
$$
(We remind the reader that the spins of the $(\beta,\gamma)$ ghosts depend
upon the value of $s$; namely $\beta$ has spin $s$ and $\gamma$ has spin
$(1-s)$.

     For the case of $s=5$, we find just one solution, one with
$\alpha^2=\ft{121}6$ given by (2.8) with $s=5$, namely $\alpha^2=\ft{121}6$.
Then  $F(\beta,\gamma,\varphi)$ in (2.3) is given by
$$
\eqalign{
F(\beta,\gamma,\varphi)&=(\del\varphi)^5 +5\alpha\, \del^2\varphi\,
(\del\varphi)^3+\ft{305}8 (\del^2\varphi)^2\, \del\varphi
+\ft{115}6 \del^3\varphi\, (\del\varphi)^2 +\ft{10}3 \alpha\,
\del^3\varphi\, \del^2\varphi\cr
&+\ft{55}{48}\alpha\, \del^4\varphi\, \del\varphi
+\ft{251}{576} \del^5\varphi +\ft{25}2 (\del\varphi)^3\,\beta\,
\del\gamma +\ft{25}4 \alpha\, \del^2\varphi\, \del\varphi \, \beta\,
\del\gamma + \ft{25}4 \alpha\, (\del\varphi)^2\, \del\beta\,\del\gamma\cr
&+\ft{125}{16} \del^3\varphi\, \beta\, \del\gamma +\ft{325}{12}
\del^2\varphi\, \del\beta\, \del\gamma +\ft{375}{16} \del\varphi\,
\del^2\beta\, \del\gamma-\ft{175}{48}\del\varphi\, \beta\, \del^3\gamma
\cr
&+\ft53 \alpha\, \del^3\beta\, \del\gamma -\ft{35}{48}\alpha\, \del\beta\,
\del^3\gamma.\cr}\eqno(2.12)
$$

     Finally, we present results for the case of $s=6$.  Here,
we find four solutions, with $\alpha^2=\ft{845}{28},\, \ft{1681}{56},\,
\ft{361}{12},\, \ft{5041}{168}$.  The first of these is the case
corresponding to the $W_6$ algebra, given by (2.8).  For this case, we
find
$$
\eqalign{
F(\beta,\gamma,\varphi)&=(\del\varphi)^6+ 6\alpha\, \del^2\varphi\,
(\del\varphi)^4 +\ft{765}7 (\del^2\varphi)^2\, (\del\varphi)^2+
\ft{256}7 \del^3\varphi\, (\del\varphi)^3 +\ft{174}{35}\alpha\,
(\del^2\varphi)^3\cr
&+\ft{528}{35}\alpha\, \del^3\varphi\, \del^2\varphi\, \del\varphi +\ft{18}7
\alpha\, \del^4\varphi\, (\del\varphi)^2 +\ft{1514}{245} (\del^3\varphi)^2
+\ft{2061}{245} \del^4\varphi\, \del^2\varphi +\ft{2736}{1225}
\del^5\varphi\, \del\varphi\cr
&+\ft{142}{6125}\alpha\, \del^6\varphi +18(\del\varphi)^4\, \beta\del\gamma +
\ft{72}{5}\alpha\, \del^2\varphi\, (\del\varphi)^2\, \beta\, \del\gamma
+\ft{48}5 \alpha\, (\del\varphi)^3\, \del\beta\, \del\gamma \cr
&+\ft{216}5
\del^3\varphi\, \del\varphi\, \beta\,\del\gamma
+\ft{1494}{35} (\del^2
\varphi)^2 \,\beta\,\del\gamma +\ft{5256}{35} \del^2\varphi\,
\del\varphi\, \del\beta\, \del\gamma+\ft{324}5 (\del\varphi)^2
\,\del^2\beta\,\del\gamma\cr
&-\ft{72}7 (\del\varphi)^2\, \beta\,\del^3\gamma
+\ft{204}{175}\alpha\, \del^4\varphi\,\beta\, \del\gamma +\ft{192}{25}\alpha\,
\del^3\varphi\, \del\beta\, \del\gamma+\ft{2376}{175}\alpha\, \del^2\varphi\,
\del^2\beta\, \del\gamma\cr
&-\ft{144}{175}\alpha\, \del^2\varphi\, \beta\, \del^3\gamma
+\ft{1296}{175}\alpha\, \del\varphi\, \del^3\beta\,
\del\gamma-\ft{576}{175}\alpha\, \del\varphi\, \del\beta\,
\del^3\gamma+\ft{1614}{175} \del^4\beta\, \del\gamma\cr
&-\ft{216}{35}
\del^2\beta\, \del^3\gamma +\ft{144}{1225}
\beta\,\del^5\gamma+\ft{144}{35} \del\beta\,\beta\,
\del^2\gamma\,\del\gamma.\cr}\eqno(2.13)
$$
We have also solved for $F(\beta,\gamma,\varphi)$ for the other values for
the background charge $\alpha$, but we shall not present them explicitly here.

\bigskip
\noindent{\bf 3. Physical states}
\bigskip

     We have seen in section 2, both from general arguments, and from
explicit solutions, that there exist nilpotent BRST operators for the
spin-2 plus spin-$s$ systems of the form (2.1--2.3).  We may now consider
solving for physical states in the corresponding string theories, by
requiring that they be annihilated by $Q_B$ but that they should not be
BRST trivial.

     We begin by considering the spin-2 plus spin-$s$ theory with the
background charge $\alpha$ given by (2.8).  In this case, the spin-2 and
spin-$s$ currents can be thought of as a subset of the spin $2,\, 3\,
\ldots,\, s$ currents of the $W_s$ algebra.  It was shown in [14] that if
one constructs the $W_s$ currents in terms of $\varphi$ and the currents
of $W_{s-1}$, then the algebra of the $W_{s-1}$ currents has the central
charge conjugate to that of the lowest non-trivial minimal model of the
$W_{s-1}$ algebra (the generalisation of the Ising model), namely
$c={2(s-2)\over (s+1)}$. Thus for our spin-2 plus spin-$s$ theory, we should
find that $T^{\rm eff}$ has central charge conjugate to that of the lowest
$W_{s-1}$ minimal model, namely
$$
c^{\rm eff}=26-{2(s-2)\over(s+1)}.\eqno(3.1)
$$
Indeed this is the case, as can be seen by substituting (2.8) into (2.9).

     The above discussion leads one to expect that the physical states of
the spin-2 plus spin-$s$ string theory should be described by a set of
Virasoro-like strings for an energy momentum tensor $T^{\rm eff}$ with
central charge given by (3.1), and intercepts given by $\Delta=1-h$, where $h$
takes values that include those of the lowest $W_{s-1}$ minimal model.  (Since
the constraints for the spin-2 plus spin-$s$ string are fewer than than in a
$W_s$ string, the physical-state conditions should be less stringent, and
therefore they should admit more solutions.)  We shall now discuss the examples
of $s=4$, 5 and 6 in detail, and see that indeed these expectations are
fullfilled.

     Consider first the spin-2 plus spin-4 string.  From (3.1), we see that
$T^{\rm eff}$ in this case has central charge $26-\ft45$, conjugate to the
lowest $W_3$ minimal model, which has $c=\ft45$.  Fortuitously in this case,
this central charge also coincides with a Virasoro minimal model, namely the
$N=5$ model (the three-state Potts model).  Thus we would expect to find
effective Virasoro strings with intercepts $\Delta=1-h$ for a set of $h$ values
that includes the conformal weights of the lowest $W_3$ minimal model, and is
included in the set of weights of the $N=5$ Virasoro minimal model.  Physical
states of ``standard'' ghost structure are described in the spin-2 plus spin-4
string by BRST-invariant operators with ghost number $G=4$ of the form
$$
U=c\, \del^2\gamma\, \del\gamma\, \gamma\, e^{\mu\varphi} e^{ip\cdot X}.
\eqno(3.2)
$$
(For convenience, we always discuss physical states that are tachyonic from the
point of view of the effective spacetime.  One can always replace the tachyon
vertex operator $e^{ip\cdot X}$ by any excited effective physical operator,
{\it i.e.}\ by an operator involving excitations of $X^\mu$ which is highest
weight under $T^{\rm eff}$, with the same intercept $\Delta$ as the tachyon.)
Acting with $Q_B$ of (2.1), with $F(\beta,\gamma,\varphi)$ given by (2.10),
and requiring that (3.2) be annihilated, we find that  the $\varphi$ momentum
$\mu$ in (3.2) must take one of the values $\mu=-\ft89\alpha,\,
-\ft{10}9\alpha,\, -\ft{26}{27}\alpha,\, -\ft{28}{27}\alpha$, and  that
correspondingly the intercept for $e^{ip\cdot X}$ takes the values
$\Delta=1,\, 1,\, \ft{14}{15},\, \ft{14}{15}$.  At level  $\ell=1$ in
excitations of $\beta$, $\gamma$ and $\varphi$, we find physical operators
with ghost number $G=3$ of the form
$$
U=c\, \del\gamma\, \gamma\, e^{\mu\varphi} e^{ip\cdot X},\eqno(3.3)
$$
with $\mu=-\ft23 \alpha,\, -\ft{16}{27}\alpha,\, -\ft{20}{27}\alpha$, and
$\Delta=\ft35,\, \ft{14}{15},\, \ft13$ respectively.  We have solved for all
the physical  operators at levels $\ell$ up to and including 9.  The results
for the  $\varphi$ momenta $\mu$, and the effective intercepts $\Delta$, are
given in  Table 1 below.
$$
\hbox{
\vbox{\tabskip=0pt \offinterlineskip
\def\tablerule{\noalign{\hrule}}
\halign to350pt{\strut#& \vrule#\tabskip=0em plus10em&
\hfil#\hfil& \vrule#& \hfil#& \vrule#& \hfil#& \vrule#&
\hfil#& \vrule#\tabskip=0pt\cr\tablerule
&&\phantom{}&&${\scriptstyle G}$&&${\scriptstyle \Delta}\ \quad$
&&$\mu\ ({\scriptstyle {\rm In\ units\ of}\ \alpha/27) \quad}$&\cr\tablerule
&&$\ell=0$
&&${\scriptstyle 4}$
&&${\scriptstyle {14\over15}\quad\ 1 }$
&&${\scriptscriptstyle \{-28,-26\}\quad \{-30,-24\}}$&\cr
\tablerule
&&$\ell=1$
&&${\scriptstyle 3}$
&&${\scriptstyle {1\over3}\quad {3\over5} \quad {14\over15}}$
&&${\scriptstyle -20\quad -18\quad -16}$&\cr
\tablerule
&&$\ell=2$
&&${\scriptstyle 3}$
&&${\scriptstyle -{2\over5}\ \quad {1\over3}}$
&&${\scriptstyle -18\quad -14}$&\cr
\tablerule
&&$\ell=3$
&&${\scriptstyle 2}$
&&${\scriptstyle {1\over3}\quad {14\over15}}$
&&${\scriptstyle -10\quad -\phantom{1}8}$&\cr
\tablerule
&&$\ell=4$
&&${\scriptstyle 2}$
&&${\scriptstyle {3\over5}}$
&&${\scriptstyle  -\phantom{1}6}$&\cr
\tablerule
&&$\ell=5$
&&${\scriptstyle 2}$
&&${\scriptstyle -{2\over5}\ \quad {1\over3}}$
&&${\scriptstyle -\phantom{1}6\quad -\phantom{1}4}$&\cr
\tablerule
&&$\ell=6$
&&${\scriptstyle 1}$
&&${\scriptstyle 1}$
&&${\scriptstyle 0}$&\cr
\tablerule
&&$\ell=7$
&&${\scriptstyle 1}$
&&${\scriptstyle {14\over15}}$
&&${\scriptstyle 2}$&\cr
\tablerule
&&$\ell=8$
&&${\scriptstyle 1}$
&&${\scriptstyle {14\over15}}$
&&${\scriptstyle 4}$&\cr
\tablerule
&&$\ell=9$
&&${\scriptstyle 1}$
&&${\scriptstyle -2\ \quad 1}$
&&${\scriptstyle 0\quad\phantom{-1}6}$&\cr
\tablerule
\noalign{}}}}
$$
\centerline{\it Table 1. Continuous-momentum physical operators for the spin-2
plus spin-4 string}

\medskip

     One can see from the results that the momentum $\mu$ in the $\varphi$
direction is always frozen to values that are integer multiples of
$\ft1{27}\alpha$.  If we assume that this is true in general,  $\mu={k\over
27} \alpha$, then it is easy to see that the mass-shell condition $-\ft12
\mu^2 -\mu \alpha +\ft12 p^2 +p\cdot a =7-\ell$, {\it i.e.}\ $-\ft12 \mu^2
-\mu \alpha +\Delta=7-\ell$, implies that a necessary condition for the
existence of a solution with effective intercept $\Delta$ at level $\ell$ is
that there exist an integer $k$ such that
$$
(k+27)^2=120(\Delta +\ell) -111.\eqno(3.4)
$$
We believe, for reasons that will emerge below, that the physical states that
we have found explicitly, and that are tabulated in Table 1, include
representatives with all the possible intercepts $\Delta$ for the spin-2 plus
spin-4 string.

     There are also discrete physical states in the theory, with zero
momentum in the effective spacetime.  One can apply (3.4), with $\Delta=0$,
to determine the levels at which such states might arise.  The result is
$\ell=1,\, 7,\, 10,\, 28,\, 34,\, \ldots$.  At $\ell=7$ there is a solution
with $k=0$; this corresponds simply to the identity operator $1$, which gives
rise to the $SL(2,C)$ vacuum as physical state.  At the next allowed level,
$\ell=10$, we can expect to find the analogue of the ground-ring operator
that arises at level 2 in the two-scalar string [17], and at level 6 in the
$W_3$ string [10].  One can associate a screening current $S$ with any physical
operator $U$, according to the prescription
$$
S(w)=\oint dz\, b(z)U(w).\eqno(3.5)
$$
It was shown in [9] that by applying this to the level 6 discrete state of the
$W_3$ string, one obtains, after dropping a total derivative term, the very
simple screening current $S=\beta\, e^{\ft27 \alpha\varphi}$. Thus we may
expect that an analogous screening current should exist here for the spin-2
plus spin-4 string.  From (3.4), it should have momentum $\mu=\ft29\alpha$,
and indeed we find that such a screening current,
$$
S=\beta\, e^{\ft29 \alpha\varphi}_\dum\eqno(3.6)
$$
exists ({\it i.e.}\ $Q_B$ acting on $S$ gives a total derivative).  Note that
$S$ has ghost number $G=-1$.  It satisfies
$$
\{Q_B,S\}=\del D,\eqno(3.7)
$$
where $D$ is the level $\ell=10$ discrete state mentioned above.

     It was argued in [12] that one can expect the screening current analogous
to (3.6) in the $W_3$ string to act as a generator of all the higher-level
physical states, by acting on the lowest-level representatives for each
effective intercept $\Delta$ with powers of the screening charge.  Similarly,
we can expect here that the charge constructed by integrating (3.6) can be
applied to the lowest-level representative in each sector of physical states
listed in Table 1 in order to generate the entire higher-level
spectrum\footnote{$^*$}{\tenfoot In fact in this case,
and presumably in general, it seems that $\Delta$ alone does not fully
characterise the sectors.  In the lowest $W_3$ minimal model, the fields with
conformal weights 0 and $\ft25$ have zero weight under the spin-3
primary current, whilst the fields with conformal weights $\ft1{15}$ and
$\ft23$ each occur twice, once with a positive, and once with an equal and
opposite negative weight under the spin-3 current.  This is reflected in our
results for the physical states, where one can see evidence for two
independent sequences of $\Delta=\ft{14}{15}$ states and two independent
sequences of $\Delta=\ft13$ states.  Within each sequence, the
$\varphi$ momenta are such that they can be obtained from the
lowest-level member by the action of the screening charge, but the
$\varphi$ momenta for the two sequences cannot be matched by any integer
powers of the screening charge.}.  In view of its $\varphi$ momentum
$\ft29\alpha$, and its ghost number $G=-1$, one expects that the trend for the
higher-level states is for the $\varphi$ momentum to increase, and the ghost
number to decrease.  This is analogous to what happens in the $W_3$ string.

     Let us now consider the relation of our results for the spin-2 plus
spin-4 string to the $c=\ft45$ minimal models discussed above.  The
conformal weights of the primary fields of the lowest $W_3$ minimal model,
which has $c=\ft45$, are $h=\{0,\, \ft1{15},\, \ft25,\, \ft23\}$ [18].  One can
see from Table 1 that these are conjugate to the effective intercepts $\Delta$
of some of the physical states, in the sense that $\Delta=1-h$.  On the other
hand, the $N=5$ Virasoro minimal model also has $c=\ft45$, and its primary
fields have dimensions $h=\{0,\, \ft1{40},\, \ft1{15},\, \ft18,\, \ft25,\,
\ft{21}{40},\, \ft23,\, \ft75,\, \ft{13}8,\, 3\}$.  We see from Table 1 that
the full set of effective intercepts $\Delta$ is conjugate to a subset of the
Virasoro minimal model weights, namely $1-\Delta=h=\{0,\, \ft1{15},\, \ft25,\,
\ft23,\, \ft75,\, 3\}$.  In fact one can easily check from the fusion rules
[19] for the $N=5$ Virasoro minimal model that this particular subset of fields
closes on itself.  This provides an indication that the set of effective
intercepts that we have found by studying levels up to $\ell=9$ in the spin-2
plus spin-4 string is probably complete.  In fact, part
of our motivation in searching to levels as high as $\ell=9$ was to find the
``missing'' intercept $\Delta=-2$, which does not make its appearance until
this level.  We expect that the three-point correlation functions of the
physical operators of the spin-2 plus spin-4 string, calculated using the
procedures introduced in [11,9], will be in agreement with the fusion rules of
the closed subset of $c=\ft45$ Virasoro minimal model fields listed above.
Higher-point correlation functions should be consistent with these three-point
functions.  The spin-2 plus spin-4 string seems to admit the interpretation of
a $c=25\ft15$ matter system with fields $X^\mu$, coupled to a $c=\ft45$ model
realised by the $\{\varphi,\beta,\gamma\}$ system.  Note that here, and for
all the spin-2 plus spin-$s$ strings that we are considering in this paper,
the physical states can be divided into ``prime states,'' occurring at the
lowest ghost number for a given level $\ell$, and higher ghost number
partners built by acting on the prime states with the ghost boosters
$a_\varphi\equiv [Q_B,\varphi]$ and $a_{X^\mu}^\dum\equiv [Q_B,X^\mu]$.  ( All
of the physical states listed in the Tables in this paper are prime states.)
It is sometimes necessary to use the ghost boosters in order to obtain
non-vanishing correlation functions (analogous to the use of picture changing
in the superstring).  The structure of the multiplets of states generated by
the ghost boosters is discussed in detail in [11,9] for the $W_3$ string; the
same considerations apply for the string theories in this paper.

     For the spin-2 plus spin-5 string, with $\alpha$ given by (2.8) for
$s=5$, we have studied physical states up to and including level $\ell=13$.
We find $\ell=0$ prime states with ``standard'' ghost structure, described by
ghost number $G=5$ physical operators of the form $U=c\, \del^3\gamma\,
\del^2\gamma\, \del\gamma\, \gamma\, e^{\mu \varphi} e^{i p\cdot X}$, with
$\mu=-\ft{10}{11}\alpha,\, -\ft{12}{11}\alpha,\, -\ft{21}{22}\alpha,\,
-\ft{23}{22}\alpha,\, -\alpha$.  The corresponding values for the effective
intercepts are $\Delta=1,\, 1,\, \ft{15}{16},\, \ft{15}{16},\, \ft{11}{12}$.
At level $\ell=1$, we find prime states corresponding to $G=4$ physical
operators of the form $U=c\,  \del^2\gamma\, \del\gamma\, \gamma\, e^{\mu
\varphi} e^{i p\cdot X}$, with $\mu=-\ft8{11},\, -\ft9{11},\, -\ft{15}{22},\,
-\ft{17}{22}$, and corresponding intercept values $\Delta= \ft23,\, \ft14,\,
\ft{15}{16},\, \ft7{16}$.  The results up to level 13 are displayed in Table 2
below.

     Again, we see that the $\varphi$ momentum is always frozen to values that
are integer multiples of a basic quantum, in this case $\mu={k\over 22}
\alpha$.  From the mass-shell condition we therefore find that a necessary
condition for the existence of a physical state with effective intercept
$\Delta$ at level $\ell$ is that there exist an integer $k$ such that
$$
(k+22)^2=48(\Delta+\ell) -44.\eqno(3.8)
$$
Discrete states with zero momentum in the effective spacetime could
therefore arise at levels $\ell=1,\, 3,\, 5,\, 11,\, 15,\, 25,\, 31,\,
\ldots$.  The $SL(2,C)$ vacuum is a discrete physical state at $\ell=11$,
corresponding to the identity operator.  The analogue of the ground-ring
operator could be expected to arise at the next permitted level, namely
$\ell=15$ with $\varphi$ momentum $\mu=\ft2{11}\alpha$.  We have checked
explicitly, and found that indeed the ghost number $G=-1$ operator
$$
S=\beta\, e^{\ft2{11}\alpha\varphi}_\dum\eqno(3.9)
$$
is a screening current, with
$$
\{Q_B,S\}=\del D,\eqno(3.10)
$$
where $D$ is the $\ell=15$ discrete state.  One should again be able to act
with powers of the screening charge, built by integrating (3.9), on the
lowest-level representative of each class of physical states, characterised
in part by $\Delta$, in order to generate all the higher-level ones.  (As for
the $s=4$ case, there can be more than one class of states with a given
$\Delta$; for example when $\Delta=\ft{15}{16}$.)

     From (3.1) we see that in this case, where $s=5$, the effective theory
has central charge 25, conjugate to the central charge $c=1$ of the lowest
$W_4$ minimal model.  Thus we should expect that a subset  of the effective
intercept values $\Delta$ should be conjugate to the weights
$$
\hbox{
\vbox{\tabskip=0pt \offinterlineskip
\def\tablerule{\noalign{\hrule}}
\halign to400pt{\strut#& \vrule#\tabskip=0em plus20em&
\hfil#\hfil& \vrule#& \hfil#& \vrule#& \hfil#& \vrule#&
\hfil#& \vrule#\tabskip=0pt\cr\tablerule
&&\phantom{}&&${\scriptstyle G}$&&${\scriptstyle \Delta}\qquad\qquad$
&&$\mu\ {\scriptstyle ({\rm In\ units\ of}\ \alpha/22)}$\qquad&\cr
\tablerule
&&$\ell=0$
&&${\scriptstyle 5}$
&&${\scriptstyle {11\over 12}}\quad\phantom{-}{\scriptstyle {15\over 16}}
\quad{\scriptstyle \phantom{1}1}$
&&${\scriptstyle -22}\quad
{\scriptscriptstyle \{-23, -21\}}\quad {\scriptscriptstyle \{-24,-20\}}$&\cr
\tablerule
&&$\ell=1$
&&${\scriptstyle 4}$
&&${\scriptstyle {1\over 4}}\quad{\scriptstyle \phantom{-}{7\over 16}
\quad {\scriptstyle \phantom{-1}{2\over 3}}
\quad {\scriptstyle {15\over16}}}$
&&${\scriptstyle -18}\quad {\scriptstyle -17}\quad {\scriptstyle -16}\quad
{\scriptstyle -15}$&\cr
\tablerule
&&$\ell=2$
&&${\scriptstyle 4}$
&&${\scriptstyle -{9\over 16}}\quad \ {\scriptstyle
-{1\over 3}}\quad{\scriptstyle \phantom{1}{1\over 4}}$
&&${\scriptstyle -17}\quad
{\scriptstyle -16}\quad {\scriptstyle -14}$&\cr
\tablerule
&&$\ell=3$
&&${\scriptstyle 3}$
&&${\scriptstyle {\scriptstyle \phantom{-}0}\, \quad\phantom{-}{7\over16}}
\quad{\scriptstyle {11\over 12}}$
&&${\scriptstyle -12}\quad {\scriptstyle -11}\quad {\scriptstyle -10}
$&\cr
\tablerule
&&$\ell=4$
&&${\scriptstyle 3}$
&&${\scriptstyle -{9\over16}}\quad {\scriptstyle {7\over16}}$
&&${\scriptstyle -11}\quad {\scriptstyle  -\phantom{1}9} $&\cr
\tablerule
&&$\ell=5$
&&${\scriptstyle 3}$
&&${\scriptstyle -2 }\quad{\scriptstyle -{13\over12} }\quad
{\scriptstyle -{9\over16}} \quad {\scriptstyle \phantom{1}0}$
&&${\scriptstyle -12} \quad {\scriptstyle -10} \quad
 {\scriptstyle -\phantom{1}9 \quad {\scriptstyle -\phantom{1}8}} $&\cr
\tablerule
&&$\ell=6$
&&${\scriptstyle 2}$
&&${\scriptstyle -{25\over12}} \quad{\scriptstyle {15\over16}}$
&&${\scriptstyle -\phantom{1}6} \quad {\scriptstyle -\phantom{1}5} $&\cr
\tablerule
&&$\ell=7$
&&${\scriptstyle 2}$
&&${\scriptstyle \phantom{-}{2\over3}}$
&&${\scriptstyle -\phantom{1}4} $&\cr
\tablerule
&&$\ell=8$
&&${\scriptstyle 2}$
&&${\scriptstyle -{1\over3}} \quad{\scriptstyle {7\over16}}$
&&${\scriptstyle -\phantom{1}4} \quad {\scriptstyle -\phantom{1}3} $&\cr
\tablerule
&&$\ell=9$
&&${\scriptstyle 2}$
&&${\scriptstyle -{33\over16}} \quad{\scriptstyle -{9\over16}}
\quad {\scriptstyle \phantom{1}{1\over4}}$
&&${\scriptstyle -\phantom{1}5}\quad {\scriptstyle -\phantom{1}3} \quad
{\scriptstyle -\phantom{1}2} $&\cr
\tablerule
&&$\ell=10$
&&${\scriptstyle 1}$
&&${\scriptstyle 1}$
&&${\scriptstyle 0}
$&\cr
\tablerule
&&$\ell=11$
&&${\scriptstyle 1}$
&&${\scriptstyle {15\over16}}$
&&${\scriptstyle 1}$&\cr
\tablerule
&&$\ell=12$
&&${\scriptstyle 1}$
&&${\scriptstyle {11\over12}}$
&&${\scriptstyle 2}$&\cr
\tablerule
&&$\ell=13$
&&${\scriptstyle 1}$
&&${\scriptstyle -2} \quad{\scriptstyle {15\over16}}$
&&${\scriptstyle 0}\quad {\scriptstyle \phantom{-1}3}$&\cr
\tablerule
\noalign{}}}}
$$
\centerline{\it Table 2. Continuous-momentum physical operators for the spin-2
plus spin-5 string}
\medskip

\noindent
of this minimal
model, namely $h=\{0,\, \ft1{16},\, \ft1{12},\, \ft13,\, \ft9{16},\, \ft34,\,
1\}$ [20].  Indeed this is the case, as can be seen by looking at the results
in Table 2.  Unlike the spin-2 plus spin-4 string discussed above, where the
central charge $c=\ft45$ of the $W_3$ minimal model happened to coincide with
that of a Virasoro minimal model, here the central charge $c=1$ of the $W_4$
minimal model does not coincide with a standard Virasoro minimal model.  On
the other hand, it is known that there do exist ``curiosities'' at $c=1$ in
the Virasoro algebra [21], and it may well be that the set of weights
$h=1-\Delta$ conjugate to the effective intercepts for the spin-2 plus spin-5
string are those of such a curiosity.  The weights that we have found, from
Table 2, are: $1-\Delta=h=\{ 0,\,\ft1{16},\,\ft1{12},\,\ft13,\,\ft9{16},\,
\ft34,\, 1,\, \ft43,\,\ft{25}{16},\, 3,\,\ft{49}{16},\, \ft{37}{12} \}$.  We
do not know if this exhausts the list, or whether further values might arise
if one looks at higher-level physical states.

     An interesting feature of the spin-2 plus spin-5 string is that the
central charge of the effective spacetime energy-momentum tensor $T^{\rm
eff}$ takes an integer value, namely $c^{\rm eff}=25$.  This means that in
this case one avoids the necessity of including a background-charge vector
$a_\mu$ in (2.7), by choosing $d=25$.  It may be that the resulting string
theory can be viewed as a 25-dimensional Virasoro string coupled to a
``curiosity at $c=1$,'' realised by the $\{\varphi,\beta,\gamma\}$ system.

     For the spin-2 plus spin-6 string, with $\alpha$ given by (2.8) for
$s=6$, we have analysed the physical states up to level $\ell=6$.  Standard
states correspond to $\ell=0$ physical operators with ghost number $G=6$, of
the form $U=c\, \del^4\gamma\, \del^3\gamma\, \del^2\gamma\, \del\gamma\,
\gamma\, e^{\mu\varphi}e^{i p\cdot X}$.  Our findings for the ghost numbers,
effective intercepts and $\varphi$ momenta for all levels up to and including
6 are presented in Table 3 below.

$$
\hbox{
\vbox{\tabskip=0pt \offinterlineskip
\def\tablerule{\noalign{\hrule}}
\halign to450pt{\strut#& \vrule#\tabskip=0em plus10em&
\hfil#\hfil& \vrule#& \hfil#& \vrule#& \hfil#& \vrule#&
\hfil#& \vrule#\tabskip=0pt\cr\tablerule
&&\phantom{}&&${\scriptstyle G}$&&${\scriptstyle \Delta}\qquad\qquad\qquad$
&&$\mu\ {\scriptstyle({\rm In\ the\ units\
of}\ \alpha/65)}$\qquad&\cr\tablerule &&$\ell=0$ &&${\scriptstyle 6}$
&&${\scriptstyle {32\over35} \quad \phantom{-}{33\over35}}
\phantom{-}\quad\,{\scriptstyle 1}$
&&${\scriptscriptstyle \{-66,-64\} \quad \{-68,-62\} \quad \{-70,
-60\}}\ $&\cr  \tablerule
&&$\ell=1$
&&${\scriptstyle 5}$
&&${\scriptstyle {1\over5}\ \quad\phantom{-} {12\over35} \quad
\phantom{-}{18\over35}
\quad\, \phantom{-}{5\over7} \quad \phantom{-}{33\over35}}$
&&${\scriptstyle -56 \quad -54 \quad -52 \quad -50 \quad
-48}\ $&\cr \tablerule
&&$\ell=2$
&&${\scriptstyle 5}$
&&${\scriptstyle -{23\over35} \quad -{17\over35}\, \quad -{2\over7}
\quad\ \phantom{-}{1\over5} }$
&&${\scriptstyle -54 \quad -52 \quad -50 \quad -46}\ $&\cr
\tablerule
&&$\ell=3$
&&${\scriptstyle 4}$
&&${\scriptstyle -{1\over5} \quad\  \phantom{-}{1\over7}
\quad \phantom{-}{18\over35} \quad \phantom{-}{32\over35}}$
&&${\scriptstyle -42 \quad -40 \quad -38 \quad -36}\ $&\cr
\tablerule
&&$\ell=4$
&&${\scriptstyle 4}$
&&${\scriptstyle -{6\over7} \quad -{17\over35} \quad \phantom{-}{12\over35}}$
&&${\scriptstyle  -40 \quad -38 \quad -34}
\ $&\cr
\tablerule
&&$\ell=5$
&&${\scriptstyle 4}$
&&${\scriptstyle -{11\over5} \quad -{13\over7} \quad -{52\over35}
\quad -{38\over35} \quad -{23\over35} \quad\  -{1\over5}}$
&&${\scriptstyle -42 \quad -40 \quad -38 \quad -36 \quad -34 \quad -32}
\ $&\cr
\tablerule
&&$\ell=6$
&&${\scriptstyle 3}$
&&${\scriptstyle -{1\over5} \quad \phantom{-}{12\over35} \quad
\phantom{-}{32\over35}}$
&&${\scriptstyle -28 \quad -26 \quad -24}
\ $&\cr
\tablerule
\noalign{}}}}
$$
\centerline{\it Table 3. Continuous-momentum physical operators for the
spin-2 plus spin-6 string}

\medskip

     We see in this case that the $\varphi$ momentum is frozen to values of
the form $\mu={k\over 65}\alpha$, where $k$ is an integer.  From the mass-shell
condition, it therefore follows that a necessary condition for the existence
of a physical state with effective intercept $\Delta$ at level $\ell$ is that
there should exist an integer $k$ such that
$$
(k+65)^2=280(\Delta+\ell)-255.\eqno(3.11)
$$
Discrete states with zero momentum in the effective spacetime might
therefore occur at levels $\ell=1,\, 16,\, 21,\, 66,\, \ldots$.  The
$SL(2,C)$ vacuum corresponds to the identity operator at $\ell=16$, and the
analogue of the ground-ring operator can be expected to arise at level
$\ell=21$, with $\varphi$ momentum $\mu=\ft2{13}\alpha$.  We have checked that
indeed the $G=-1$ current
$$
S=\beta\, e^{\ft2{13}\alpha\varphi}_\dum\eqno(3.12)
$$
is a screening current, satisfying
$$
\{Q_B,S\}=\del D,\eqno(3.13)
$$
with $D$ being the level $\ell=21$ discrete state.  Again, the screening
charge obtained by integrating (3.12) can be expected to generate
higher-level physical states from the lowest-level representatives for each
effective intercept $\Delta$.  (As in the lower-$s$ examples, there can be
more than one class of states for a given $\Delta$.  This occurs, for
example, in the $\Delta=\ft{32}{35}$ and $\Delta=\ft{33}{35}$ sectors.)

     The central charge for the effective energy-momentum tensor $T^{\rm eff}$
is, from (3.1), equal to $26-\ft87$ for this spin-2 plus spin-6 string.  Thus
it is conjugate to the central charge $c=\ft87$ of the lowest $W_5$ minimal
model.  This model has primary fields with conformal weights $h=\{0,\,
\ft2{35},\, \ft3{35},\, \ft27,\, \ft{17}{35},\, \ft{23}{35},\, \ft45,\,
\ft67,\, \ft65\}$ [20].  From the list of effective intercepts $\Delta$ in
Table 3, we see that the conjugate weights $h=1-\Delta$ include all of the
$W_5$ minimal-model weights given above, together with  $h=\{\ft97,\,
\ft{13}7,\, \ft{20}7,\, \ft{16}5,\, \ft{52}{35},\, \ft{58}{35},\,
\ft{73}{35},\, \ft{87}{35}\}$.  It may well be that there are also further
effective intercept values that will arise only at levels beyond $\ell=6$.
Presumably the full set of conjugate conformal weights $h$ will be associated
with some particular $c=\ft87$ Virasoro model, as realised by the $\{
\varphi,\beta,\gamma\}$ system.  Little is known about such $c>1$ models.

\bigskip
\noindent{\bf 4. Unitarity, and the other BRST operators}
\bigskip

     In section 3 we concentrated on the spin-2 plus spin-$s$ BRST operators
for which the background-charge parameter $\alpha$ is given by (2.8).  These
cases presumably correspond to truncations of the $W_s$ algebra in which all
but the spin-2 and spin-$s$ currents are omitted.  It seems that the
corresponding string theories are likely to be unitary, in the sense that
the effective intercept values $\Delta$ will not give rise to non-unitary
states of the effective Virasoro string theory described by $T^{\rm eff}$
with central charge given by (3.1).  For example, one can derive the
following limits on the intercept values for a Virasoro string with central
charge $c$, by requiring that level-1 and level-2 excited states have no
negative-norm degrees of freedom:
$$
\eqalignno{
\underline{\rm level \ 1}:\qquad& \Delta\le 1,&(4.1a)\cr
\underline{\rm level \ 2}:\qquad&
\Delta \le {\textstyle{37-c-\sqrt{(c-1)(c-25)}\over 16}},\quad {\rm or}\quad
 \Delta \ge {\textstyle{37-c+\sqrt{(c-1)(c-25)}\over 16}}.&(4.1b)\cr}
$$
For the case of the spin-2 plus spin-4  string, this means that unitarity
requires that $\Delta\le \ft35$ or $\ft78  \le \Delta\le 1$.  From the results
in Table 1, we see that these conditions  are satisfied by all the sectors of
the theory.  Whilst this does not  constitute a complete proof of unitarity,
it certainly provides a strong  indication that it holds for the spin-2 plus
spin-4 theory.  For the spin-2  plus spin-5 string, where $c^{\rm eff}=25$,
the conditions (4.1$a$--$b$) reduce to  $\Delta\le 1$, which is satisfied by
all the sectors of physical states  listed in Table 2.  For the spin-2 plus
spin-6 string, and indeed for all  cases with $s\ge6$, the value of $c^{\rm
eff}$ is less than 25.  Under these  circumstances, there is no unitarity
restriction coming from level-2 excited  states.  The fact that all the
sectors of physical states given in Table 3  for the spin-2 plus spin-6 string
have $\Delta\le1$ indicates that this  theory is probably unitary too.
Presumably this persists for all the higher spin-2  plus spin-$s$ strings, in
the case that $\alpha$ is given by (2.8).

     The story may be different for the solutions for spin-2 plus spin-$s$
BRST  operators corresponding to other values of $\alpha$ that are not given
by (2.8).   For example, we found a second spin-2 plus spin-4 BRST operator,
with $Q_1$  given by (2.3) and (2.11), with $\alpha^2=\ft{361}{30}$.  Solving
for physical  states, we find that there are standard ghost-structure states
at $\ell=0$  with $\mu=-\ft{17}{19},\, -\ft{21}{19},\, -\ft{18}{19},\,
-\ft{20}{19}$ and  effective intercepts $\Delta=\ft{21}{20},\, \ft{21}{20},\,
1,\, 1$  respectively.  We have checked all levels up to and including
$\ell=6$, and  we find the following set of effective intercepts:
$$
\Delta=\{\ft{21}{20},\, 1,\, \ft45,\, \ft14 \}.\eqno(4.2)
$$
The fact that one of the intercept values exceeds 1 is suggestive of
non-unitarity of the physical states in this sector of the effective string
theory.  In fact, $c^{\rm eff}=26\ft35 >26$, and so it would perhaps not be
surprising if there were some difficulty with unitarity in this case.
However, it is worth noting that the value of the intercept that exceeds 1,
namely $\ft{21}{20}$, coincides with the limit of the second inequality
in (4.1$b$); in other words, there would be a null state in the spectrum of
physical states with level 2 excitations in the effective spacetime.  It may
be that a 2-scalar model, {\it i.e.}\ $\varphi$ together with just one extra
scalar $X$, would make sense even in a case where intercepts greater than 1
occur.

     Similar remarks apply to the other BRST operators that we have found for
the spin-2 plus spin-6 system.  The case $\alpha^2=\ft{361}{12}$ is
potentially interesting because it corresponds to $c^{\rm eff}=26$.  We have
looked at the $\ell=0$ physical states, and found the effective intercepts
$\Delta=\{\ft{26}{25},\, 1,\, \ft{24}{25}\}$. At $\ell=1$, we find
$\Delta=\{1,\, \ft{21}{25},\, \ft{14}{25},\, \ft{11}{25},\, \ft4{25}\}$.  For
$\alpha^2=\ft{1681}{56}$, we find $\ell=0$ intercepts $\Delta=\{\ft{15}{14},\,
\ft{115}{112},\, 1\}$, and for $\alpha^2=\ft{5041}{168}$, we find $\ell=0$
intercepts   $\Delta=\{\ft{15}{14},\, \ft{117}{112},\, 1\}$.

\bigskip
\noindent{\bf 5. Conclusions}
\bigskip

     We have given the general form for BRST operators for a class of
extensions of the Virasoro algebra, where there is a primary current of spin
$s$ in addition to the energy-momentum tensor.  We have found the explicit
forms for the BRST operators in the cases $s=4$, 5 and 6.  The case $s=3$
corresponds to the well-known $W_3$ algebra, which has been well studied over
the last few years.

     One may use these BRST operators to build anomaly-free extensions of
two-dimensional gravity, and hence to build extensions of ordinary string
theory.  We have discussed the detailed structure of the physical states for
the cases $s=4$, 5 and 6.  For the class of spin-2 plus spin-$s$ BRST operators
for which the background charge $\alpha$ for $\varphi$ is given by (2.8), the
effective spacetime energy-momentum tensor $T^{\rm eff}$ has the a central
charge $c^{\rm eff}$ that is conjugate to that of the lowest $W_{s-1}$ minimal
model, in the sense that $26=c^{\rm eff} + c^{\rm min}$.  One expects
therefore that the effective intercept values $\Delta$ for the physical states
should be conjugate to a set of weights $h$ that at least include
those of the associated $W_{s-1}$ minimal model, in the sense that $1=\Delta
+ h$.  This is indeed what we have found, in the explicit examples of $s=4$,
5 and 6.  However, because the physical states are subject to constraints
only from a spin-2 and a spin-$s$ current, there are more sectors than just
those that are conjugate to the primary fields of the $W_{s-1}$ minimal model.
When $s=4$, for which $c^{\rm eff}=26-\ft45$, we find intercept values
conjugate to the weights of a subset of the fields of the 3-state Potts model
that close under the fusion rules.  For $s=5$, where $c^{\rm eff}=26-1$, we
find intercepts conjugate to the weights of the primary fields of a $c=1$
Virasoro model.  For $s=6$, and higher, the intercepts are conjugate to models
with $c>1$.  The spin-2 plus spin-$s$ BRST operators with the background
charge $\alpha$ given by (2.8) all appear to give rise to string theories that
are unitary in the effective spacetime.

    For the examples of spin-2 plus spin-4, and spin-2 plus spin-6 BRST
operators, we have also found other solutions for which the central
charge is not given by (2.8).  Such solutions presumably exist for higher
values of $s$ too.  These BRST operators appear to give rise to string
theories that are not unitary from the effective spacetime point of view;
in particular, one finds sectors of the theory with effective intercepts
that exceed 1.

     In a sense, all of the $W$-string theories that have been constructed
suffer from the drawback that they can be reinterpreted as special cases of
ordinary Virasoro strings in which one builds a $c=26$ energy-momentum tensor
as the direct sum of a spacetime energy-momentum tensor $T^{\rm eff}$ and an
energy-momentum tensor $T^{\rm min}$ for a minimal model or some other model
containing a finite number of primary fields.  On the other hand, one can
take the view that this close connection between minimal models and $W$
strings reveals an interesting underlying $W$ symmetry as an organising
principle for the physical states.

     For the $W_3$ string it is known that the structure of physical states
in the two-scalar realisation is richer than in the multi-scalar realisations
analogous to those that we have been considering in this paper.
Specifically, one finds that the physical states of the multi-scalar $W_3$
string can all be understood as generalisations of some of the two-scalar
physical states, but that the two-scalar theory also has further physical
states that do not generalise beyond two dimensions [10,11,9].  A similar
phenomenon occurs for the two-scalar spin-2 plus spin-$s$ strings that we have
been considering in this paper.  For example, we find
that there is an $\ell=3$ physical operator in the two-scalar spin-2 plus
spin-4 string, with $\alpha^2=\ft{243}{20}$, of the form
$$
U=\Big( c\, \gamma + \ft43 \del^2\gamma\, \gamma -\ft{16}9 \alpha\,
\del\varphi\, \del \gamma\, \gamma -\ft8{11}a \, \del X\, \del \gamma\,
\gamma -\ft85 b\, c\, \del\gamma\, \gamma \Big) e^{-\ft23 \alpha \varphi
+\ft6{11} a X},\eqno(5.1)
$$
where $a$ is the background charge for the single extra field $X$, and
$a^2=\ft{121}{60}$. This operator does not generalise to the multi-scalar
case, when $X$ is replaced by $X^\mu$.  It may be that just as for the $W_3$
string, the more subtle aspects of the underlying $W$ algebra are better
captured by the two-scalar than the multi-scalar realisations.

\np
\centerline{\bf ACKNOWLEDGMENTS}
\bigskip

     We are grateful to Stany Schrans for discussions.  We have made extensive
use of the Mathematica package OPEdefs [22], written by Kris Thielemans, for
the computations in this paper.

\bigskip
\singlespace
\centerline{\bf REFERENCES}
\frenchspacing
\bigskip

\item{[1]}A.B. Zamolodchikov, {\sl Teor. Mat. Fiz.} {\bf 65} (1985)
1205.

\item{[2]}C.M. Hull, {\sl Phys. Lett} {\bf B240} (1989) 110.

\item{[3]}A. Bilal and J.-L. Gervais, {\sl Nucl. Phys.} {\bf B314} (1989)
646.

\item{[4]}S.R. Das, A. Dhar and S.K. Rama, {\sl Mod. Phys. Lett.}
{\bf A6} (1991) 3055; {\sl Int. J. Mod. Phys.} {\bf A7} (1992) 2295.

\item{[5]}C.N. Pope, L.J. Romans and K.S. Stelle, {\sl Phys.
Lett.} {\bf B268} (1991) 167; {\sl Phys. Lett.} {\bf B269} (1991) 287;\nl
C.N. Pope, L.J. Romans, E. Sezgin and K.S. Stelle,
{\sl Phys. Lett.} {\bf B274} (1992) 298.

\item{[6]}J. Thierry-Mieg, {\sl Phys. Lett.} {\bf B197} (1987) 368.

\item{[7]}F. Bais, P. Bouwknegt, M. Surridge and K. Schoutens, {\sl Nucl.
Phys} {\bf B304} (1988) 348.

\item{[8]}L.J.  Romans, {\sl Nucl.  Phys.} {\bf B352} (1991) 829.

\item{[9]}H. Lu, C.N. Pope, S. Schrans and X.J. Wang,  ``On the spectrum
and scattering of $W_3$ strings,'' preprint CTP TAMU-4/93, KUL-TF-93/2,
hep-th/9301099.

\item{[10]}C.N. Pope, E. Sezgin, K.S. Stelle and X.J. Wang, {\sl Phys. Lett.}
{\bf B299} (1993) 247.

\item{[11]}H. Lu, C.N. Pope, S. Schrans and X.J. Wang,  ``The interacting
$W_3$ string,'' preprint CTP TAMU-86/92, KUL-TF-92/43, hep-th/9212117, to
appear in {\sl Nucl. Phys.} {\bf B}.

\item{[12]}M.D. Freeman and P.C. West, ``The covariant scattering and
cohomology of $W_3$ strings,'' preprint, KCL-TH-93-2, hep-th/9302114.

\item{[13]}M.D. Freeman and P.C. West, ``$W_3$ string scattering,''
preprint, KCL-TH-92-4, hep-th/9210134.

\item{[14]}H. Lu, C.N. Pope, S. Schrans and K.W. Xu,  {\sl Nucl.
Phys.} {\bf B385} (1992) 99.

\item{[15]}E. Bergshoeff, H.J. Boonstra, M. de Roo, S. Panda and A. Sevrin,
``On the BRST operator of $W$ strings,'' preprint, UG-2/93.

\item{[16]}E. Bergshoeff, H.J. Boonstra, M. de Roo, S. Panda, A. Sevrin and
X. Shen, to appear.

\item{[17]}E. Witten, {\sl Nucl. Phys.} {\bf B373} (1992) 187;\nl
E. Witten and B. Zwiebach, {\sl Nucl. Phys.} {\bf B377} (1992) 55.

\item{[18]}V.A. Fateev and A.B. Zamolodchikov, {\sl Nucl. Phys.} {\bf B280}
(1987) 644.

\item{[19]}A.A. Belavin, A.M. Polyakov and A.B. Zamolodchikov, {\sl Nucl.
Phys.} {\bf B241} (1984) 333.

\item{[20]}V.A. Fateev and A.B. Zamalodchikov, {\sl Int. J. Mod. Phys.} {\bf
A3} (1988) 507.

\item{[21]}P. Ginsparg, {\sl Nucl. Phys.} {\bf B295} (1988) 153.

\item{[22]}K. Thielemans, {\sl Int. J. Mod. Phys.} {\bf C2} (1991) 787.

\end